\documentclass{endm}
\usepackage{endmmacro}

\usepackage{pstricks,pst-node} 
\usepackage{equationarray} 

\begin{document}
\newtheorem{property}{Property} 

\begin{verbatim}\end{verbatim}\vspace{2.5cm} 

\begin{frontmatter}

\title{Approximability of the Multiple Stack TSP}

\author{Sophie Toulouse\thanksref{stemail}}
\address{LIPN (UMR CNRS 7030) - Institut Galil\'ee - Universit\'e Paris 13,\\99 av. Jean-Baptiste Cl\'ement, 93430 Villetaneuse, France}

\thanks[stemail]{Email: \href{mailto:sophie.toulouse@lipn.univ-paris13.fr}{\texttt{\normalshape sophie.toulouse@lipn.univ-paris13.fr}}} 

\begin{abstract}
STSP seeks a pair of pickup and delivery tours in two distinct networks, where the two tours are related by LIFO contraints. We address here the problem approximability. We notably establish that  asymmetric MaxSTSP and MinSTSP12 are $\mathbf{APX}$, and propose a heuristic that yields to a 1/2, 3/4 and 3/2 standard approximation for respectively Max2STSP, Max2STSP12 and Min2STSP12.
\end{abstract}


\end{frontmatter}

\section{Introduction}\label{sec-intro}
The {\em multiple Stack Traveling Salesman Probem}, STSP, has been recently introduced in~\cite{PM09} and is very well motivated by the initial request of a transportation company. Consider a company that owns a vehicle fleet and a depot in $m$ distinct cities $C_1,\ldots,C_m$. For any pair $(C_j,C_{j'})$ of cities, the company handles orders from $C_j$ suppliers to $C_{j'}$ customers, proceeding as follows: in $C_j$, a single vehicle picks up all the orders that involve a $C_{j'}$ customer; during this tour, the commodities are packed within a single container that consists of a given number $k$ of rows; the container is then sent to $C_{j'}$, where a local vehicle delivers the commodities. The operator thus is faced with two levels of transportation: the local one inside the cities, the long-haul one between the cities. STSP models the local level problem; namely, in the context we have just described, every pair of cities defines an instance of STSP. The problem specificity relies on the way the containers are managed: their rows behave like stacks, and {\em no repacking} is allowed. The rows of the containers thus are subject to LIFO {\em ``Last In First Out''} constraints. Therefore, the delivery tour in $C_{j'}$ must handle {\em at first} orders that have been handled {\em at last} in $C_{j}$. 

Formally, an instance $I$ of STSP considers a number $k$ of stacks (the container rows), together with two distinct networks $I^A=(G^A,d^A),I^B=(G^B,d^B)$, where $V(G^{\alpha})\equiv[n+1]$ and $d^{\alpha}:E(G^{\alpha})\rightarrow\mathbb{N}$, $\alpha\in\{A,B\}$. In both networks, vertex $0$ represents the depot, whereas vertex $i\in[n]$ represents the $i$th order (the $i$th supplier in $I^A$, the $i$th customer in $I^B$). A solution consists of a pickup tour $\vec{T}^A$ on $I^A$, a delivery tour $\vec{T}^B$ on $I^B$, and a packing $\mathcal{P}=\{P^1,\ldots,P^k\}$ of the commodities into the stacks, where a stack $P^\beta$ is described as a sequence $(i_1,\ldots,i_{p^{\beta}})$ of commodities (index $1$ refers to the bottom of $P^\beta$). We denote by $<^\alpha, \alpha\in\{A,B\}$ the complete order that $\vec{T}^\alpha$ induces on $[n]$: $i<^\alpha i'$ {\em iff} $\vec{T}^\alpha$ handles order $i$ earlier than it handles $i'$. Similary, we denote by $<^{\mathcal{P}}$ the partial order that $\mathcal{P}$ induces on $[n]$: $i <^{\mathcal{P}} i'$ {\em iff} $\mathcal{P}$ stacks $i$ at a lower position than $i'$ in a same stack $P^\beta$. The LIFO constraints may thus be expressed in terms of the three orders $<^A$, $<^B$, $<^{\mathcal{P}}$. Namely, a triple $(\mathcal{P},\vec{T}^A,\vec{T}^B)$ is consistent {\em iff} it satisfies:
\begin{equation}\label{eq-feasibility}
	\forall i\neq i'\in [n],\ i <^{\mathcal{P}} i' \Rightarrow (i <^A i') \wedge (i >^B i')
\end{equation}

The value of a solution $(\mathcal{P},\vec{T}^A,\vec{T}^B)$ is given by the sum $d^A(\vec{T}^A)+d^B(\vec{T}^B)$ of the distance of its two tours. The ultimate goal naturally consists in finding the feasible triple $(\mathcal{P},\vec{T}^A,\vec{T}^B)$ of optimum value.

In what follows, the stacks are assumed to have {\em unlimited capacities}. The most often, input graphs are complete. For TSP ({\em the Traveling Salesman Problem}), as well as for STSP, A(S)TSP, $\Delta$(S)TSP and (S)TSP$ab$ respectively refer the asymmetric, the metric and the bivaluated cases. Finally, $k$STSP considers that $k$ is a universal constant. 
Since one manipulates both maximization and minimization goals, we use notations $\succeq$, $\succ$, $\mathop{opt}$, $\mathop{\overline{opt}}$ instead of $\geq$, $>$, $\max$, $\min$ ({\em resp.}, $\leq$, $<$, $\min$, $\max$) if the goal is to maximize ({\em resp.}, to minimize). 
Finally, because one jointly considers pickup and delivery tours, we respectively denote by $E^{-1}=\{(i',i)\ |\ (i,i')\in E\}$, $G^{-1}=(V,E^{-1})$, $d^{-1}:E^{-1}\rightarrow\mathbb{N}:d^{-1}(i,i')=d(i',i)$ and $I^{-1}=(G^{-1},d^{-1})$ the reverse of an edge set $E$, a graph $G=(V,E)$, a distance $d:E\rightarrow\mathbb{N}$, a network $I=(G,d)$. Hence, $\vec{T}$ is a tour of value $d(\vec{T})$ on $I$ {\em iff} $\vec{T}^{-1}$ is a tour of value $d(\vec{T})$ on $I^{-1}$.

\section{Preliminaries}\label{sec-prem}
The problem trivially is $\mathbf{NP-hard}$, from TSP. Indeed, for extremal values of the number of stacks, STSP strictly is equivalent to TSP: if STSP assumes a single stack, then it reduces to TSP, considering the distance $d:(i,i')\mapsto d^A(i,i')+d^B(i',i)$; if on the opposite STSP assumes $n$ stacks, then it reduces to two independent TSP (packings that store one commodity per stack do not induce any LIFO constraint). The litterature on computational aspects of STSP is rather thin: see \cite{PM09,FOT09} for metaheuristic approachs, \cite{LLER09}~for an exact approach (that solves 2STSP via the computation of the $k$ best tours in $I^A,I^B$), \cite{TWC09} for an analysis of STSP complexity. We here make use of natural connections to TSP and observations made in~\cite{TWC09} in order to locate STSP within the approximation hierarchy.

Besides its practical impacts, STSP presents a rather intricate combinatorial structure. One the one hand, one could wonder on what part of the problem does structure / impact the most the solution / its complexity. It appeared in \cite{TWC09} that both the packing and the tour subproblems are tractable:
\begin{thm}[\cite{TWC09}]\label{thm-cpx}In arbitrary input graphs:
\begin{enumerate}
\item Deciding, given a pair $(\vec{T}^A,\vec{T}^B)$, whether it admits or not a consistent packing $\mathcal{P}$  (and to design this packing if the answer is YES) is in $\mathbf{P}$.\label{enum-cpx-P}
\item {\bf When $k$ is a universal constant:} given a packing $\mathcal{P}$, computing  a pair of tours that are optimal for $\mathcal{P}$ is in $\mathbf{P}$.
\label{enum-cpx-T}
\end{enumerate}
\end{thm} 

On the other hand, STSP has strong connections to {\em coloring} and {\em scheduling}. Indeed, subproblem~(\ref{enum-cpx-P}) reduces to graph coloring in $C(\vec{T}^A,\vec{T}^B)=\{ i\neq i'\in[n]\ |\ i<^A i' \Leftrightarrow i<^B i'\}$, whereas subproblem~(\ref{enum-cpx-T}) reduces to scheduling under $<^{\mathcal{P}}$ (or $>^{\mathcal{P}}$) precedence constraints. 

\section{A first classification of STSP}\label{sec-apx}
\begin{property}\label{pty-reduc}
	\begin{enumerate}\label{enum-reduction}
	\item\label{enum-reduction-stsp-tsp} There exists a standard approximation preserving reduction from ASTSP to ATSP that maps a $\rho$-approximate solution of ATSP to a $\rho/2$, $\rho/2+(a/2b)$, $\rho/2+(b/2a)$-approximate solution of ASTSP for Max(S)TSP, Max(S)TSP$ab$ and Min(S)TSP$ab$, respectively.
	\item\label{enum-reduction-tsp-stsp} {\em Case of arbitrary graphs.} Any approximate factor $\rho$ for any restriction of STSP also holds for the same restricton of TSP.
	\end{enumerate}	
\end{property}
\begin{proof}
\ref{enum-reduction-stsp-tsp}. Associate with the instance $I=(I^A,I^B)$ of ASTSP the pair $(I^A,I^B)$ of instances of ATSP. Compute $\vec{T}^\alpha, \alpha\in\{A,B\}$ an approximate tour on $I^\alpha$. Pick the tour $\vec{T}^A$ or $\vec{T}^B$ that performs the best among $d^A(\vec{T}^A)+(d^B)^{-1}(\vec{T}^A)$, $(d^A)^{-1}(\vec{T}^B)+d^A(\vec{T}^B)$.
\ref{enum-reduction-tsp-stsp}. Associate with the instance $I=(G,d)$ of TSP the instance $I'=(I,I^{-1})$ of STSP.
\end{proof}

\begin{proposition}[Immediate from Property~\ref{pty-reduc}]\label{propo-classification}
~\\(i)~Max(A)STSP, Max$\Delta$(A)STSP, Max(A)STSP01, \\Min(A)STSP12 are standard-approximable within the following  factor:\\
\begin{tabular}{cccc}
		MaxSTSP						&MaxASTSP				&Max$\Delta$STSP				&Max$\Delta$ASTSP\\[-9pt]
		$61/162-o(1)$, \cite{COW05}	&$1/3$, \cite{KLSS05}	&$7/16-o(1)$, \cite{CN07}		&$31/80$, \cite{BSS09}\\
		MaxSTSP01					&MaxASTSP01				&MinSTSP12						&MinASTSP12\\[-9pt]
		$3/7$, \cite{BK08}			&$3/8$, \cite{B04}		&11/7, \cite{BK08}				&13/8, \cite{B04}\\
\end{tabular}
\\(ii)~ Max$\Delta$STSP$\in\mathbf{MaxSNP}$-hard, \cite{BJWW98}; 
					MaxSTSP12$\in\mathbf{APX}$-hard, \cite{PY93}; 
					MinSTSP is not $2^{p(n+1)}$ standard-approximable for any polynomial $p$, or $\mathbf{P}=\mathbf{NP}$ %
\end{proposition}

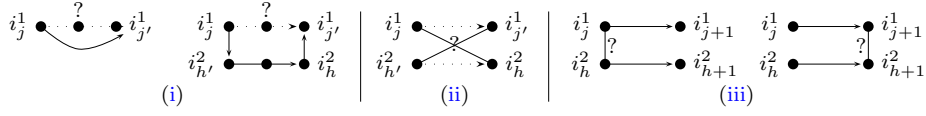
\begin{figure}[t]
\begin{center}
\begin{pspicture}(0.2,0.2)(11,1.0)
\uput[-90](1.75,0.4){\scriptsize{(\ref{lem-partialsol-1})}}
\qdisk(0,1.0){2pt}\uput[180](0.0,1.0){\scriptsize{$i^1_j$}}
\qdisk(0.5,1.0){2pt}\uput[90](0.5,0.95){\scriptsize{?}}
\qdisk(1.0,1.0){2pt}\uput[0](1.0,1.0){\scriptsize{$i^1_{j'}$}}
\pcline[linewidth=0.4pt,linestyle=dotted]{->}(0.0,1.0)(1.1,1.0)
\pscurve[linewidth=0.2pt]{->}(0.0,1.0)(0.5,0.7)(1.1,0.9)
\qdisk(2.5,1.0){2pt}\uput[180](2.5,1.0){\scriptsize{$i^1_j$}}
\qdisk(3.0,1.0){2pt}\uput[90](3.0,0.95){\scriptsize{?}}
\qdisk(3.5,1.0){2pt}\uput[0](3.5,1.0){\scriptsize{$i^1_{j'}$}}
\qdisk(3.5,0.5){2pt}\uput[0](3.5,0.5){\scriptsize{$i^2_{h}$}}
\qdisk(3.0,0.5){2pt}
\qdisk(2.5,0.5){2pt}\uput[180](2.5,0.5){\scriptsize{$i^2_{h'}$}}
\pcline[linewidth=0.4pt,linestyle=dotted]{->}(2.5,1)(3.4,1)
\pcline[linewidth=0.2pt]{->}(2.5,0.5)(3.4,0.5)
\pcline[linewidth=0.2pt]{->}(2.5,1.0)(2.5,0.6)
\pcline[linewidth=0.2pt]{->}(3.5,0.5)(3.5,0.9)
\pcline[linewidth=0.2pt]{-}(4.25,0.0)(4.25,1.2)
\uput[-90](5.5,0.4){\scriptsize{(\ref{lem-partialsol-2})}}
\qdisk(5.0,1.0){2pt}\uput[180](5.0,1.0){\scriptsize{$i^1_j$}}
\qdisk(6.0,1.0){2pt}\uput[0](6.0,1.0){\scriptsize{$i^1_{j'}$}}
\qdisk(6.0,0.5){2pt}\uput[0](6.0,0.5){\scriptsize{$i^2_{h}$}}
\qdisk(5.0,0.5){2pt}\uput[180](5.0,0.5){\scriptsize{$i^2_{h'}$}}
\pcline[linewidth=0.4pt,linestyle=dotted]{->}(5.0,1.0)(5.9,1.0)
\pcline[linewidth=0.4pt,linestyle=dotted]{->}(5.0,0.5)(5.9,0.5)
\pcline[linewidth=0.2pt]{-}(5.0,1.0)(6.0,0.5)
\pcline[linewidth=0.2pt]{-}(6.0,1.0)(5.0,0.5)\uput[90](5.5,0.5){\scriptsize{?}}
\pcline[linewidth=0.2pt]{-}(6.75,0.0)(6.75,1.2)
\uput[-90](9.25,0.4){\scriptsize{(\ref{lem-partialsol-3})}}
\qdisk(7.5,1.0){2pt}\uput[180](7.5,1.0){\scriptsize{$i^1_j$}}
\qdisk(8.5,1.0){2pt}\uput[0](8.5,1.0){\scriptsize{$i^1_{j+1}$}}
\qdisk(8.5,0.5){2pt}\uput[0](8.5,0.5){\scriptsize{$i^2_{h+1}$}}
\qdisk(7.5,0.5){2pt}\uput[180](7.5,0.5){\scriptsize{$i^2_{h}$}}
\pcline[linewidth=0.2pt]{->}(7.5,1.0)(8.4,1.0)
\pcline[linewidth=0.2pt]{->}(7.5,0.5)(8.4,0.5)
\pcline[linewidth=0.2pt]{-}(7.5,1.0)(7.5,0.5)\uput[0](7.35,0.75){\scriptsize{?}}
\qdisk(10.0,1.0){2pt}\uput[180](10.0,1.0){\scriptsize{$i^1_j$}}
\qdisk(11.0,1.0){2pt}\uput[0](11.0,1.0){\scriptsize{$i^1_{j+1}$}}
\qdisk(11.0,0.5){2pt}\uput[0](11.0,0.5){\scriptsize{$i^2_{h+1}$}}
\qdisk(10.0,0.5){2pt}\uput[180](10.0,0.5){\scriptsize{$i^2_{h}$}}
\pcline[linewidth=0.2pt]{->}(10.0,1.0)(10.9,1.0)
\pcline[linewidth=0.2pt]{->}(10.0,0.5)(10.9,0.5)
\pcline[linewidth=0.2pt]{-}(11.0,1.0)(11.0,0.5)\uput[180](11.15,0.75){\scriptsize{?}}
\end{pspicture}
\caption{\label{fig-partialsol}\small{The three configurations that contradict consistency.}}
\end{center}
\end{figure}

\section{A matching based heuristic for the symmetric 2STSP}\label{sec-2stsp}
In what follows, we say that an edge set $N$ is {\em consistent} with a packing $\mathcal{P}$ {\em iff} there exists a completion ${N'}$ of $N$ into a tour $T$ {\em s.t.} $T$ is feasible with respect to $\mathcal{P}$. 
We first provide a characterization of consistency for 2STSP.
\begin{proposition}[Due to lack of space, the proof is omitted.]\label{lem-partialsol}
Given a packing $\mathcal{P}$, $J^{\beta}=\{0,\ldots,p^{\beta}+1\}$ denotes the index set on $P^\beta$, $\beta\in\{1,2\}$. Artificial indexes $0,p^{\beta}+1$ are introduced in order to represent the depot vertex, {\em i.e.}, $i^\beta_0=i^\beta_{p^\beta+1}=0$. Let $N$ be a collection of pairwise disjoint elementary chains on $[n+1]$, then $N$ is consistent with $\mathcal{P}$ {\em iff} it satifies the three following properties (configurations that contradict consistency are illustrated in Figure~\ref{fig-partialsol}):
\begin{enumerate}
	\item\label{lem-partialsol-1}No jump: $\forall\beta\in\{1,2\},\ \forall j<j'\in J^{\beta}$,
		$\left(\{i^{\beta}_j,i^{\beta}_{j'}\}\in N\right)$\\
				\hspace*{1cm}$\vee\ \left(\exists h\leq h'\in J^{3-\beta},\ \{i^{\beta}_j,i^{3-\beta}_{h},\ldots,i^{3-\beta}_{h'},i^{\beta}_{j'}\}\subseteq N\right) \ \Rightarrow\ j'=j+1$
	\item\label{lem-partialsol-2}No crossing edges: $\forall j\neq j'\in J^1,\ \forall h\neq h'\in J^2$, 
		$\{\{i^1_j,i^2_{h}\},\{i^1_{j'},i^2_{h'}\}\}\subseteq N$\\
				\hspace*{1cm}$\Rightarrow\ (j<j' \Leftrightarrow h<h')$ 
	\item\label{lem-partialsol-3}No way back: $\forall j\in J^1,\ \forall h\in J^2,$
		$\{\{i^1_j,i^1_{j+1}\},\{i^2_h,i^2_{h+1}\}\}\subseteq N$\\
				\hspace*{1cm}$\Rightarrow\ \left(\{i^1_j,i^2_h\}\right) \notin N \wedge \left(\{i^1_{j+1},i^2_{h+1}\}\right) \notin N$
\end{enumerate}
\end{proposition}

\begin{figure}[t]
\begin{center}
\begin{pspicture}(-0.1,-0.1)(11.5,1.2)
\uput[-90](1.20,0.1){\scriptsize{$j,j'\neq 1,\ j\equiv j'+1[2]$}}
\qdisk(0.0,0.75){2pt}\uput[90](0.0,0.7){\scriptsize{$0$}}
\qdisk(0.5,1.0){2pt}\uput[90](0.5,1.0){\scriptsize{$c^1_2$}}
\qdisk(1.0,1.0){2pt}\uput[90](1.0,1.0){\scriptsize{$c^1_3$}}
\qdisk(1.5,1.0){2pt}\uput[90](1.5,1.0){\scriptsize{$c^1_4$}}
\qdisk(2.0,1.0){2pt}\uput[90](2.0,1.0){\scriptsize{$c^1_5$}}
\qdisk(2.5,1.0){2pt}\uput[90](2.5,1.0){\scriptsize{$c^1_6$}}
\qdisk(1.5,0.5){2pt}\uput[-90](1.5,0.5){\scriptsize{$c^1_7$}}
\qdisk(1.0,0.5){2pt}\uput[-90](1.0,0.5){\scriptsize{$c^1_8$}}
\qdisk(0.5,0.5){2pt}\uput[-90](0.5,0.5){\scriptsize{$c^1_9$}}
\pcline[linewidth=0.2pt]{-}(0.0,0.75)(0.5,1.0)
\pcline[linewidth=0.4pt,linestyle=dotted]{-}(0.5,1.0)(1.0,1.0)
\pcline[linewidth=0.2pt]{-}(1.0,1.0)(1.5,1.0)
\pcline[linewidth=0.4pt,linestyle=dotted]{-}(1.5,1.0)(2.0,1.0)
\pcline[linewidth=0.2pt]{-}(2.0,1.0)(2.5,1.0)
\pcline[linewidth=0.4pt,linestyle=dotted]{-}(0.0,0.75)(0.5,0.5)
\pcline[linewidth=0.2pt]{-}(0.5,0.5)(1.0,0.5)
\pcline[linewidth=0.4pt,linestyle=dotted]{-}(1.0,0.5)(1.5,0.5)
\pcline[linewidth=0.2pt,linestyle=dashed]{-}(2.0,1.0)(1.0,0.5)\uput[0](1.55,0.75){\scriptsize{$e^*$}}
\uput[-90](4.60,0.1){\scriptsize{$j,j'\neq 1,\ j\equiv j'[2]$}}
\qdisk(3.1,0.75){2pt}\uput[90](3.1,0.7){\scriptsize{$0$}}
\qdisk(3.6,1.0){2pt}\uput[90](3.6,1.0){\scriptsize{$c^1_2$}}
\qdisk(4.1,1.0){2pt}\uput[90](4.1,1.0){\scriptsize{$c^1_3$}}
\qdisk(4.6,1.0){2pt}\uput[90](4.6,1.0){\scriptsize{$c^1_4$}}
\qdisk(5.1,1.0){2pt}\uput[90](5.1,1.0){\scriptsize{$c^1_5$}}
\qdisk(5.6,1.0){2pt}\uput[90](5.6,1.0){\scriptsize{$c^1_6$}}
\qdisk(4.6,0.5){2pt}\uput[-90](4.6,0.5){\scriptsize{$c^1_9$}}
\qdisk(4.1,0.5){2pt}\uput[-90](4.1,0.5){\scriptsize{$c^1_8$}}
\qdisk(3.6,0.5){2pt}\uput[-90](3.6,0.5){\scriptsize{$c^1_7$}}
\qdisk(6.1,0.75){2pt}\uput[-90](6.1,0.8){\scriptsize{$0$}}
\pcline[linewidth=0.1pt]{-}(3.1,0.75)(3.6,1.0)
\pcline[linewidth=0.4pt,linestyle=dotted]{-}(3.6,1.0)(4.1,1.0)
\pcline[linewidth=0.1pt]{-}(4.1,1.0)(4.6,1.0)
\pcline[linewidth=0.4pt,linestyle=dotted]{-}(4.6,1.0)(5.1,1.0)
\pcline[linewidth=0.1pt]{-}(5.1,1.0)(5.6,1.0)
\pcline[linewidth=0.4pt,linestyle=dotted]{-}(6.1,0.75)(4.6,0.5)
\pcline[linewidth=0.1pt]{-}(4.6,0.5)(4.1,0.5)
\pcline[linewidth=0.4pt,linestyle=dotted]{-}(4.1,0.5)(3.6,0.5)
\pcline[linewidth=0.1pt,linestyle=dashed]{-}(4.6,1.0)(4.1,0.5)\uput[0](4.3,0.75){\scriptsize{$e^*$}}
\uput[-90](7.8,0.1){\scriptsize{$j'=1$}}
\qdisk(6.8,0.75){2pt}\uput[90](6.8,0.7){\scriptsize{$0$}}
\qdisk(7.3,1.0){2pt}\uput[90](7.3,1.0){\scriptsize{$c^1_2$}}
\qdisk(7.8,1.0){2pt}\uput[90](7.8,1.0){\scriptsize{$c^1_3$}}
\qdisk(8.3,1.0){2pt}\uput[90](8.3,1.0){\scriptsize{$c^1_4$}}
\qdisk(9.3,0.5){2pt}\uput[-90](9.3,0.5){\scriptsize{$c^1_5$}}
\qdisk(8.8,0.5){2pt}\uput[-90](8.8,0.5){\scriptsize{$c^1_6$}}
\qdisk(8.3,0.5){2pt}\uput[-90](8.3,0.5){\scriptsize{$c^1_7$}}
\qdisk(7.8,0.5){2pt}\uput[-90](7.8,0.5){\scriptsize{$c^1_8$}}
\qdisk(7.3,0.5){2pt}\uput[-90](7.3,0.5){\scriptsize{$c^1_9$}}
\qdisk(9.8,0.75){2pt}\uput[90](9.8,0.7){\scriptsize{$0$}}
\pcline[linewidth=0.2pt]{-}(6.8,0.75)(7.3,1.0)
\pcline[linewidth=0.4pt,linestyle=dotted]{-}(7.3,1.0)(7.8,1.0)
\pcline[linewidth=0.2pt]{-}(7.8,1.0)(8.3,1.0)
\pcline[linewidth=0.4pt,linestyle=dotted]{-}(8.3,1.0)(9.3,0.5)
\pcline[linewidth=0.2pt]{-}(9.3,0.5)(8.8,0.5)
\pcline[linewidth=0.4pt,linestyle=dotted]{-}(6.8,0.75)(7.3,0.5)
\pcline[linewidth=0.2pt]{-}(7.3,0.5)(7.8,0.5)
\pcline[linewidth=0.4pt,linestyle=dotted]{-}(7.8,0.5)(8.3,0.5)
\pcline[linewidth=0.2pt,linestyle=dashed]{-}(8.3,1.0)(9.8,0.75)\uput[45](8.95,0.8){\scriptsize{$e^*$}}
\pcline[linewidth=0.2pt]{-}(10.5,0.95)(11.0,0.95)\uput[0](11.0,0.95){\scriptsize{$M^A$}}
\pcline[linewidth=0.4pt,linestyle=dotted]{-}(10.5,0.55)(11.0,0.55)\uput[0](11.0,0.55){\scriptsize{$M^B$}}
\end{pspicture}
\caption{\label{fig-alg-neven-p1}\small{The approximate packing when $n$ is even and $p=1$.}}
\end{center}
\end{figure}
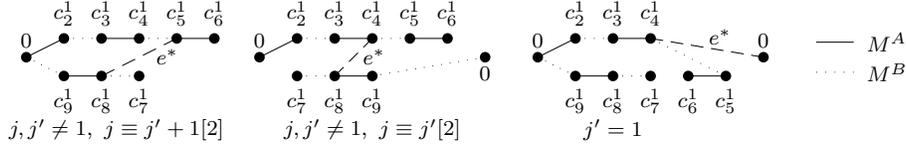

The algorithm relies on a pair $(M^A,M^B)$ of optimum weight matchings on $I^A,I^B$. Given such a pair, together with some additional edge $e^*$, it aims at computing a packing $\mathcal{P}$ {\em s.t.} $N^\alpha, \alpha\in\{A,B\}$, is consistent with $\mathcal{P}$, where $N^\alpha$ coincides with $M^\alpha$ ({\em resp.}, $M^\alpha\cup\{e^*\}$) when $n$ is odd ({\em resp.}, even). 
Let $H$ denote the multigraph $([n+1],M^A\cup M^B)$, and $\mathcal{C}^1,\ldots,\mathcal{C}^p$ denote its connected components. 
We assume {\em w.l.o.g.}, that the depot vertex ({\em i.e.}, node $0$) coincides with vertex $c^1_1$ in $\mathcal{C}^{1}$.
If $n$ is odd ({\em iff} $n+1$ is even), then every component $\mathcal{C}^h$ is an elementary cycle $\{c^h_1,\ldots,c^h_{q^h},c^h_1\}$ of even length $q^h\geq 2$. 
Otherwise, $\mathcal{C}^{h}=\{c^{h}_1,\ldots,c^{h}_{q^{h}},c^{h}_1\}\backslash\{\{c^{h}_\ell,c^{h}_{\ell+1}\}\}$ is an elementary chain of even length $q^{h}-1\geq 0$ for a lonely index $h\in[p]$. 
We define the additional edge $e^*$ as follows: 
\begin{equationarray}{lccl}
	\forall p,		&e^*	&=	&\arg\mathop{opt}\{\mathop{opt}_{\alpha=A,B}\{d^\alpha(f)\}\ |\ f\in F(p)\}\textit{, where:}\label{eq-e-star}\\
					&F(1)	&=	&\left\{\left(\{0\}\cup\{c^1_3,\ldots,c^1_{\ell}\}\right)\times
									\left(\{0\}\cup\{c^1_{\ell+1},\ldots,c^1_{n}\}\right)\right\}\backslash\{\{0,0\}\}\label{eq-F-p1}\\
					&F(p)	&=	&\cup_{(h<h')=1}^p V(\mathcal{C}^h)\times V(\mathcal{C}^{h'})\label{eq-F-pgeq2}
\end{equationarray}
The heuristic is described in Algorithm~\ref{algo-2stsp}: it first computes $M^\alpha,\ \alpha\in\{A,B\}$ (Step~(\ref{algo-2stsp-cpl})) and $e^*$ (case $n$ even, Step~(\ref{algo-2stsp-cc-e})); it then builds the approximate packing $\mathcal{P}$ by considering the components the one after each other, starting with $\mathcal{C}^1$ (Step~(\ref{algo-2stsp-P})); it finally computes the best tours $T^\alpha(\mathcal{P}), \alpha\in\{A,B\}$, with respect to the $\mathcal{P}$ (Step~(\ref{algo-2stsp-T})). Note that before it operates the packing, the algorithm may have in Step~(\ref{algo-2stsp-e-cc}) to perform some reindexation of the $\mathcal{C}^h$, depending on $e^*$ (notably, if $p\geq 2$, then the two connected components that are incident to $e^*$ must be stacked consecutively). The even case is illustrated in Figures~\ref{fig-alg-neven} ($p\geq 2$) and \ref{fig-alg-neven-p1} ($p=1$).

\begin{algorithm}
\caption{\label{algo-2stsp}\small{An appproximate algorithm for 2STSP.}}
\small{
\begin{enumerate}
	\item\label{algo-2stsp-cpl} Compute $M^\alpha$ an optimum weight matching on $I^\alpha,\alpha\in\{A,B\}$
	\item\label{algo-2stsp-cc-e} Determine the connected components of $H$ in such a way that $c^1_1=0$\\
								{\bf If} $n$ is even {\bf Then} compute $e^*=\arg\mathop{opt}\{\mathop{opt}_{\alpha=A,B}\{d^\alpha(f)\}\ |\ f\in F(p)\}$
 	\item\label{algo-2stsp-e-cc}{\bf If} $n$ is even and $p\geq 2$ {\bf Then} consider the components in such a way that:
		\begin{enumerate}
		\item 	{\bf If} $V(e^*)\cap V(\mathcal{C}^1)\in\{\emptyset,\{0\}\}$ 
						{\bf Then} $\exists h\in[p]\backslash\{1\}$ {\em s.t.} $e^*=\{c^h_{\lceil q^h/2\rceil},c^{h+1}_1\}$\\ 
			  	{\bf Else} $e^*=\{c^1_j,c^2_{1}\}$\label{algo-2stsp-e-cc-1}
				{\em // $h$ is taken modulo $[p]$, $j\in\{2,\ldots,q^1\}$ depends on $c^1_1$}
		\item 	{\bf If} $q^2\neq2\wedge e^*=\{c^1_j,c^2_{1}\}\wedge \exists \alpha, \{\{0,c^1_2\},\{c^2_1,c^2_{q^2}\}\}\subseteq M^\alpha$  
						{\bf Then} reverse $\mathcal{C}^2$\label{algo-2stsp-e-cc-2}
		\end{enumerate}
	\item\label{algo-2stsp-P} {\bf If $n$ is odd or $p\geq 2$:}\\
								\hspace*{0.3cm}$j^1\leftarrow$ ({\bf If} $n\equiv 0[2]$ $\wedge$ $e^*=\{c^1_j,c^2_1\}$ 
															{\bf Then} $j$: $\lceil (q^1-1)/2\rceil+1$ {\bf otherwise})\\
								\hspace*{0.3cm}$j^2\leftarrow$ ({\bf If} $n\equiv 0[2]$ $\wedge$ $e^*=\{c^1_2,c^2_1\}$ $\wedge$ $q^2=2$ 
															{\bf Then} $q^2$: $\lceil q^2/2\rceil$ {\bf otherwise})\\
								\hspace*{0.3cm}$P^1\leftarrow(c^1_2,\ldots,c^1_{j^1},\ \ \ \  
												c^2_1,\ldots,c^2_{j^2},\ \ \ \ 
												c^3_1,\ldots,c^3_{\lceil q^3/2\rceil}\ \ \ \ \ ,\ldots,\  
												c^p_1,\ldots,c^p_{\lceil q^p/2\rceil})$\\
								\hspace*{0.3cm}$P^2\leftarrow(c^1_{q^1},\ldots,c^1_{j^1+1}, 
												c^2_{q^2},\ldots,c^2_{j^2+1}, 
												c^3_{q^3},\ldots,c^3_{\lceil q^3/2\rceil+1}\ ,\ldots,\ 
												c^p_{q^p},\ldots,c^p_{\lceil q^p/2\rceil+1})$\\
								{\bf Else}
									(thus $p=1$, $e^*=\{c^1_j,c^1_{j'}\}$, $j\in\{1\}\cup\{3,\ldots,\ell\},j'\in\{1\}\cup\{\ell+1,\ldots,n+1\}$){\bf :}\\
								\hspace*{0.3cm}{\bf If} $j,j'\neq 1$ and $j\equiv j'+1[2]$ 
												{\bf Then} $P^1\leftarrow(c^1_2,\ldots,c^1_\ell)$, $P^2\leftarrow(c^1_{n+1},\ldots,c^1_{\ell+1})$\\
								\hspace*{0.3cm}{\bf Else If} $j,j'\neq 1$ and $j\equiv j'[2]$ 
												{\bf Then} $P^1\leftarrow(c^1_2,\ldots,c^1_\ell)$, $P^2\leftarrow(c^1_{\ell+1},\ldots,c^1_{n+1})$\\
								\hspace*{0.3cm}{\bf Else If} $j'=1$ 
												{\bf Then} $P^1\leftarrow(c^1_2,\ldots,c^1_j)$, $P^2\leftarrow(c^1_{n+1},\ldots,c^1_{j+1})$\\
								\hspace*{0.3cm}{\bf Else If} $j=1$ 
												{\bf Then} $P^1\leftarrow(c^1_2,\ldots,c^1_{j'-1})$, $P^2\leftarrow(c^1_{n+1},\ldots,c^1_{j'})$
\item\label{algo-2stsp-T}Compute $T^\alpha(\mathcal{P}), \alpha\in\{A,B\}$, the best feasible tour for $\mathcal{P}=(P^1,P^2)$ on $I^\alpha$
\end{enumerate}}
\end{algorithm}


\begin{figure}[t]
\begin{center}
\begin{pspicture}(0,0.3)(11.5,1.2)
\uput[180](0,1.25){\scriptsize{$P^1:$}}
\uput[180](0,0.25){\scriptsize{$P^2:$}}
\qdisk(0.25,0.75){2pt}\uput[180](0.30,0.75){\scriptsize{$c^1_1=0$}}
\qdisk(1.0,1.0){2pt}\uput[90](1.0,0.95){\scriptsize{$c^1_2$}}
\qdisk(1.0,0.5){2pt}\uput[-90](1.0,0.55){\scriptsize{$c^1_3$}}
\pcline[linewidth=0.4pt]{-}(0.25,0.75)(1,1.0) 	
\pcline[linewidth=0.2pt,linestyle=dotted]{-}(1.0,0.5)(0.25,0.75) 				
\qdisk(2.0,1.0){2pt}\uput[90](2.0,0.95){\scriptsize{$c^2_1$}}
\qdisk(3.0,1.0){2pt}\uput[90](3.0,0.95){\scriptsize{$c^2_2$}}
\qdisk(3.0,0.5){2pt}\uput[-90](3.0,0.55){\scriptsize{$c^2_3$}}
\qdisk(2.0,0.5){2pt}\uput[-90](2.0,0.55){\scriptsize{$c^2_4$}}
\pcline[linewidth=0.2pt]{-}(2.0,1.0)(3.0,1.0) 					
\pcline[linewidth=0.4pt,linestyle=dotted]{-}(3.0,1.0)(3.0,0.5) 	
\pcline[linewidth=0.2pt]{-}(2.0,0.5)(3.0,0.5) 					
\pcline[linewidth=0.4pt,linestyle=dotted]{-}(2.0,0.5)(2.0,1.0) 	
\qdisk(4.0,1.0){2pt}\uput[90](4.0,0.95){\scriptsize{$c^3_1$}}
\qdisk(4.0,0.5){2pt}\uput[-90](4.0,0.55){\scriptsize{$c^3_2$}}
\pcline[linewidth=0.4pt,linestyle=dotted]{-}(3.95,1.0)(3.95,0.5) 	
\pcline[linewidth=0.2pt]{-}(4.05,0.5)(4.05,1.0) 					
\qdisk(5.0,1.0){2pt}\uput[90](5.0,0.95){\scriptsize{$c^4_1$}}
\qdisk(6.0,1.0){2pt}\uput[90](6.0,0.95){\scriptsize{$c^4_2$}}
\qdisk(7.0,1.0){2pt}\uput[90](7.0,0.95){\scriptsize{$c^4_3$}}
\qdisk(7.0,0.5){2pt}\uput[-90](7.0,0.55){\scriptsize{$c^4_4$}}
\qdisk(6.0,0.5){2pt}\uput[-90](6.0,0.55){\scriptsize{$c^4_5$}}
\qdisk(5.0,0.5){2pt}\uput[-90](5.0,0.55){\scriptsize{$c^4_6$}}
\pcline[linewidth=0.4pt,linestyle=dotted]{-}(5.0,1.0)(6.0,1.0) 	
\pcline[linewidth=0.2pt]{-}(6.0,1.0)(7.0,1.0) 					
\pcline[linewidth=0.4pt,linestyle=dotted]{-}(7.0,1.0)(7.0,0.5) 	
\pcline[linewidth=0.2pt]{-}(7.0,0.5)(6.0,0.5) 					
\pcline[linewidth=0.4pt,linestyle=dotted]{-}(6.0,0.5)(5.0,0.5) 	
\pcline[linewidth=0.2pt]{-}(5.0,0.5)(5.0,1.0) 					
\uput[0](10.3,1.2){\scriptsize{$M^A$}}\pcline[linewidth=0.2pt]{-}(9.7,1.2)(10.2,1.2)
\uput[0](10.3,0.75){\scriptsize{$T^A,T^B\backslash M^\alpha$}}\pcline[linewidth=0.2pt,linestyle=dashed]{-}(9.7,0.75)(10.2,0.75)
\uput[0](10.3,0.3){\scriptsize{$M^B$}}\pcline[linewidth=0.4pt,linestyle=dotted]{-}(9.7,0.3)(10.2,0.3)
\pcline[linewidth=0.2pt, linestyle=dashed]{-}(1.0,1.0)(2.0,1.0)\uput[90](1.5,0.85){\scriptsize{$e^*$}}
\pcline[linewidth=0.2pt, linestyle=dashed]{-}(1.0,0.5)(1.0,1.0)\uput[0](0.60,0.70){\scriptsize{$^B$}}
\pcline[linewidth=0.2pt, linestyle=dashed]{-}(1.0,0.5)(3.0,1.0)\uput[0](1.25,0.70){\scriptsize{$^A$}}
\pcline[linewidth=0.2pt, linestyle=dashed]{-}(1.0,0.5)(2.0,0.5)\uput[-90](1.5,0.60){\scriptsize{$^A$}}
\pcline[linewidth=0.2pt, linestyle=dashed]{-}(2.0,0.5)(3.0,1.0)\uput[0](2.35,0.60){\scriptsize{$^B$}}
\pcline[linewidth=0.2pt, linestyle=dashed]{-}(3.0,0.5)(4.0,1.0)
\pcline[linewidth=0.2pt, linestyle=dashed]{-}(4.0,0.5)(5.0,1.0)
\pcline[linewidth=0.2pt, linestyle=dashed]{-}(5.0,0.5)(6.0,1.0)
\pcline[linewidth=0.2pt, linestyle=dashed]{-}(6.0,0.5)(7.0,1.0)
\qdisk(8.0,0.75){2pt}\uput[0](8.0,0.75){\scriptsize{$0$}}
\pcline[linewidth=0.2pt, linestyle=dashed]{-}(7.0,0.5)(8.0,0.75)
\end{pspicture}
\caption{\label{fig-alg-neven}\small{Possible completions of the approximate packing $\mathcal{P}$: case $n$ even.}}
\end{center}
\end{figure}
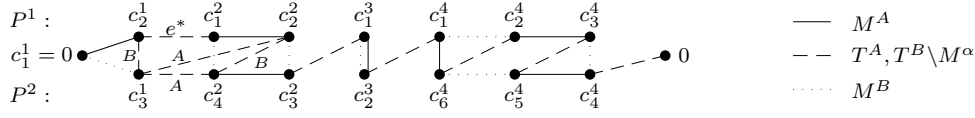

\begin{claim}\label{claim-algo}
\begin{enumerate}
\item Algorithm~\ref{algo-2stsp} has polynomial-time complexity.\label{claim-algo-cpx}
\item $N^\alpha, \alpha\in\{A,B\}$, is consistent with the approximate packing $\mathcal{P}$.\label{claim-algo-right}
\item The solutions returned by the algorithm are $3/2$, $3/4$ and $1/2$ standard approximate for Min2STSP12, Max2STSP12 and Max2STSP, respectively; these ratios are (asymptotically) tight.\label{claim-algo-apx}
\end{enumerate}
\end{claim}

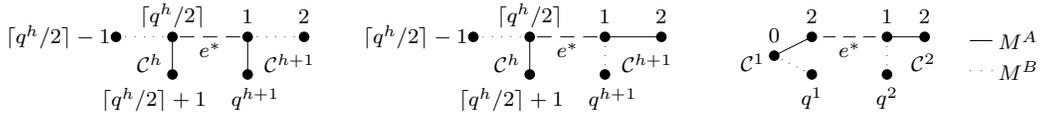
\begin{figure}[t]
\begin{center}
\begin{pspicture}(-0.3,0.2)(14,1.2)
\uput[225](2.0,0.9){\scriptsize{$\mathcal{C}^{h}$}}
\qdisk(1.25,1.0){2pt}\uput[180](1.35,1.0){\scriptsize{$\lceil q^{h}/2\rceil-1$}}
\qdisk(2.0,1.0){2pt}\uput[90](2.0,0.95){\scriptsize{$\lceil q^{h}/2\rceil$}}
\qdisk(2.0,0.5){2pt}\uput[-90](1.75,0.5){\scriptsize{$\lceil q^{h}/2\rceil+1$}}
\pcline[linewidth=0.4pt,linestyle=dotted]{-}(1.25,1.0)(2.0,1.0)
\pcline[linewidth=0.2pt]{-}(2.0,1.0)(2.0,0.5)
\pcline[linewidth=0.2pt,linestyle=dashed]{-}(2.1,1)(2.9,1)\uput[-90](2.5,1.1){\scriptsize{$e^*$}}
\uput[-45](3.1,0.9){\scriptsize{$\mathcal{C}^{h+1}$}}
\qdisk(3.0,1.0){2pt}\uput[90](3.0,1.0){\scriptsize{$1$}}
\qdisk(3.75,1.0){2pt}\uput[90](3.75,1.0){\scriptsize{$2$}}
\qdisk(3.0,0.5){2pt}\uput[-90](3.1,0.5){\scriptsize{$q^{h+1}$}}
\pcline[linewidth=0.4pt,linestyle=dotted]{-}(3.75,1.0)(3.0,1.0)
\pcline[linewidth=0.2pt]{-}(3.0,1.0)(3.0,0.5)
\uput[225](6.75,0.9){\scriptsize{$\mathcal{C}^{h}$}}
\qdisk(6,1.0){2pt}\uput[180](6.1,1.0){\scriptsize{$\lceil q^{h}/2\rceil-1$}}
\qdisk(6.75,1.0){2pt}\uput[90](6.75,0.95){\scriptsize{$\lceil q^{h}/2\rceil$}}
\qdisk(6.75,0.5){2pt}\uput[-90](6.50,0.5){\scriptsize{$\lceil q^{h}/2\rceil+1$}}
\pcline[linewidth=0.4pt,linestyle=dotted]{-}(6,1.0)(6.75,1.0)
\pcline[linewidth=0.2pt]{-}(6.75,1.0)(6.75,0.5)
\pcline[linewidth=0.2pt,linestyle=dashed]{-}(6.85,1)(7.65,1)\uput[-90](7.25,1.1){\scriptsize{$e^*$}}
\uput[-45](7.85,0.9){\scriptsize{$\mathcal{C}^{h+1}$}}
\qdisk(7.75,1.0){2pt}\uput[90](7.75,1.0){\scriptsize{$1$}}
\qdisk(8.5,1.0){2pt}\uput[90](8.5,1.0){\scriptsize{$2$}}
\qdisk(7.75,0.5){2pt}\uput[-90](7.85,0.5){\scriptsize{$q^{h+1}$}}
\pcline[linewidth=0.2pt]{-}(8.5,1.0)(7.75,1.0)
\pcline[linewidth=0.4pt,linestyle=dotted]{-}(7.75,1.0)(7.75,0.5)
\uput[225](10,0.9){\scriptsize{$\mathcal{C}^1$}}
\qdisk(10,0.75){2pt}\uput[90](10,0.75){\scriptsize{$0$}}
\qdisk(10.5,1.0){2pt}\uput[90](10.5,1.0){\scriptsize{$2$}}
\qdisk(10.5,0.5){2pt}\uput[-90](10.5,0.5){\scriptsize{$q^1$}}
\pcline[linewidth=0.2pt]{-}(10.,0.75)(10.5,1.0) 
\pcline[linewidth=0.4pt,linestyle=dotted]{-}(10,0.75)(10.5,0.5) 
\pcline[linewidth=0.2pt,linestyle=dashed]{-}(10.65,1)(11.35,1)\uput[-90](11,1.1){\scriptsize{$e^*$}} 
\uput[-45](11.7,0.9){\scriptsize{$\mathcal{C}^2$}}
\qdisk(11.5,1){2pt}\uput[90](11.5,1.0){\scriptsize{$1$}}
\qdisk(12.0,1){2pt}\uput[90](12.0,1.0){\scriptsize{$2$}}
\qdisk(11.5,0.5){2pt}\uput[-90](11.5,0.5){\scriptsize{$q^2$}}
\pcline[linewidth=0.2pt]{-}(11.5,1)(12,1) 
\pcline[linewidth=0.4pt,linestyle=dotted]{-}(11.5,1)(11.5,0.5) 
\pcline[linewidth=0.2pt]{-}(12.6,0.95)(12.9,0.95)\uput[0](12.8,0.95){\scriptsize{$M^A$}}
\pcline[linewidth=0.4pt,linestyle=dotted]{-}(12.6,0.55)(12.9,0.55)\uput[0](12.8,0.55){\scriptsize{$M^B$}}
\end{pspicture}
\caption{\label{fig-chain-ok}\small{Case $n$ even and $p\geq 2$: edge $e^*$ induces {\em ``no jump''} in $M^\alpha$.}}
\end{center}
\end{figure}

\begin{proof}
{\bf Claim~(\ref{claim-algo-cpx}).} Step~(\ref{algo-2stsp-T}) only requires a $\mathcal{O}((n+1)^2)$-time, \cite{TWC09}; for the other steps, the fact is either famous, or trivial.
{\bf Claim~(\ref{claim-algo-right}).} $N^\alpha, \alpha\in\{A,B\}$ trivially satisfies Properties~(\ref{lem-partialsol-2}) and~(\ref{lem-partialsol-3}) ({\em resp.}, (\ref{lem-partialsol-1})) of Propositon~\ref{lem-partialsol}, in any case ({\em resp.}, when $n$ is odd or $p=1$). Thus assume $n\equiv 0[2]$ and  $p\geq 2$. Steps~(\ref{algo-2stsp-e-cc}-\ref{algo-2stsp-e-cc-1}) and (\ref{algo-2stsp-P}) of the algorithm indicate: $\exists h\in[p], \exists j\in[q^h]$ {\em s.t.} $e^*=\{c^h_j,c^{h+1}_{1}\}$ and $e^*\in P^1\times (P^1\cup\{0\})$. $M^\alpha$ possibly links $c^h_j$ ({\em resp.}, $c^{h+1}_1$) to vertices (when they exist) $c^h_{j-1},c^h_{j+1}$ ({\em resp.}, $c^{h+1}_1,c^{h+1}_{q^{h+1}}$), with $c^h_{j-1}\in P^1$ (unless $e=\{c^1_2,c^2_1\}$), $c^h_{j+1}\in P^2$, $c^{h+1}_2\in P^1$ and $c^{h+1}_{q^{h+1}}\in P^2$ (unless $e^*=\{c^1_2,c^2_1\}$ and $q^2=2$). Moreover, Steps~(\ref{algo-2stsp-e-cc}-\ref{algo-2stsp-e-cc-2}) and (\ref{algo-2stsp-P}) ensure that $\mathcal{P}$ satisfies: $\neg (\{0,c^1_2\}\in M^\alpha \wedge \{c^2_1,c^2_{q^2}\}\in M^\alpha)\wedge (c^2_1,c^2_{q^2})\in P^1\times P^2)$. Hence, the chain $\{c^h_{j-1},c^h_j,c^{h+1}_1,c^{h+1}_{q^{h+1}}\}$ ({\em resp.}, $\{c^h_{j+1},c^h_j,c^{h+1}_1,c^{h+1}_{2}\}$) may never jump $c^h_{j+1}$ ({\em resp.}, $c^{h+1}_{q^{h+1}}$) in $P^2$ (see Figure~\ref{fig-chain-ok} for some illustration).
{\bf Claim~(\ref{claim-algo-apx}).}
Let $(\mathcal{P}^*,T^{A,*},T^{B,*})$ ({\em resp.}, $(\mathcal{P},T^{A},T^{B})$) denote an optimum ({\em resp.}, the approximate) solution, we denote by $OPT$ {\em resp.}, $APX$) its value; we observe:
\begin{enumerate}\label{enum-facts}
\item\label{enum-facts-solref}$\forall \alpha$ in $\{A,B\}$, $\exists$ $N'^{\alpha}$ {\em s.t.} ${T'}^\alpha = N^{\alpha}\cup {N'}^{\alpha}$ is a tour, and ${T'}^\alpha$ is consistent with $\mathcal{P}$. By optimality of $T^\alpha$ with respect to $\mathcal{P}$, we thus have: \\
	$d^\alpha(T^\alpha)\succeq d^\alpha(M^{\alpha})+d^\alpha({N'}^{\alpha})+1_{n\equiv 0[2]}d^\alpha(e^*), \alpha\in\{A,B\}$.
\item\label{enum-facts-opt-odd}If $n$ is odd, then any tour on $[n+1]$ is the union of two perfect matchings on $[n+1]$. By optimality of $M^\alpha$, we thus have:
	$2d^{\alpha}(M^{\alpha})\succeq d^\alpha(T^{\alpha,*})$.
\item\label{enum-facts-opt-even}If $n$ is even, then $\forall$ $T$ tour on $[n+1]$, $\forall e\in T$, $T\backslash\{e\}$ is the union of two almost perfect perfect matchings on $[n+1]$. Furthemore, by connectivity of $T^{\alpha,*}$, $\exists e^\alpha \in T^{\alpha,*}\cap F(p), \alpha\in\{A,B\}$. By optimality of $M^\alpha$, we thus have:
	$2d^{\alpha}(M^{\alpha})\succeq d^\alpha(T^{\alpha,*}) - d^{\alpha}(e^{\alpha})$.
\end{enumerate}

From (\ref{enum-facts-solref}), (\ref{enum-facts-opt-odd}), (\ref{enum-facts-opt-even}), we deduce:
\begin{equation}\label{eq-pvapx-gen}
            2APX \succeq OPT +2\sum_{\alpha} d^{\alpha}(N'^{\alpha}) 
                + 1_{n\equiv 0[2]}(2\sum_{\alpha}d^{\alpha}(e^*)-\sum_{\alpha}d^\alpha(e^\alpha))
\end{equation}

What enables to conclude for Max2STSP, when $n$ is odd. When $n$ is even, just observe that $e^*$ is optimal on $F(p)$, whereas $e^\alpha\in F(p),\ \alpha\in\{A,B\}$:
\begin{equation}\label{eq-pvapx-gen}
\label{eq-pvapx-gen-edge}	\sum_{\alpha}d^\alpha(e^*) \succeq \mathop{opt}_{\alpha}\{d^\alpha(e^*)\} 
          \succeq \mathop{opt}_{\alpha}\{d^\alpha(e^\alpha)\} \succeq 1/2\sum_{\alpha}d^\alpha(e^\alpha)
\end{equation}

For the bivalued case, any edge set $S$ satisfies $|S|\leq d^{\alpha}(S)\leq 2|S|, \alpha\in\{A,B\}$. When $n$ is odd, thus consider $|T^{\alpha,*}|=n+1=|N'^\alpha|$ and conclude: 
\begin{equation}\label{eq-biv-odd}
	1/2OPT \leq 2(n+1)	\leq	2\sum_{\alpha} d^{\alpha}(N'^{\alpha})	\leq 4(n+1) \leq 2OPT
\end{equation}

When $n$ is even, $|N'^\alpha|=n$. Let $Q$ denote the quantity $2(d^A(e^*)+d^B(e^*))-(d^A(e^{A})+d^B(e^{B}))$: if the goal is to maximize, assume that $Q<2$; then $d^A(e^*)+d^B(e^*)=2$ and $d^A(e^{A})+d^B(e^{B})\geq 3$, what contradicts the fact that $e^*$ is optimal. Similary, $Q$ may not reach values $5,6$ when minimizing. 

Tight instances $I_{n,a,b}$ are defined on $K_{n+1}, n\geq 1$, as: $d^\alpha(u,v)=a$ 
	if $(v=u+2 \wedge u\equiv 1[4])$ or $(v=u+2 \wedge u\equiv 0[4]\wedge\alpha=A)$ or $(v=u+2 \wedge u\equiv 2[4]\wedge\alpha=B)$ 
	or $(v=u+1 \wedge u\equiv 1[2]\wedge\alpha=A)$ or $(v=u+1\wedge u\equiv 0[2]\wedge\alpha=B)$, $d^\alpha(u,v)=b$ otherwise. 
We consider subfamilies $I_{n,a,b}$ {\em s.t.} $(a,b)=(1,0)$ ({\em resp.}, $(2,1)$, $(1,2)$) for Max{2}STSP ({\em resp.}, Max{2}STSP12, Min{2}STSP12) and $n=8q-1$ ({\em resp.}, $8q$) for the odd ({\em resp.}, even) case.
\end{proof}

\begin{thm}[Immediate form Claim~\ref{claim-algo}]\label{thm-2stsp}
Min2STSP12, Max2STSP12 and Max2STSP respectively are $3/2$, $3/4$ and $1/2$ standard approximable.
\end{thm}

\section{Conclusion}\label{sec-conc}
Two main questions arise from the proposed analysis: Min$\Delta$STP approximability, STSP differential approximability.

\end{document}